# A Unified Phase-native Computational Principle Governs Hippocampal Spike Timing and Neural Coding

Reza Ahmadvand, Sara Safura Sharif, Yaser Mike Banad[*]

School of Electrical and Computer Engineering, University of Oklahoma, Oklahoma, United States.

iamrezaahmadvand1@ou.edu, s.sh@ou.edu, bana@ou.edu

[*]Corresponding author

**Abstract**

Hippocampal neurons exhibit precise phase locking to network oscillations, but the computational principle governing this temporal precision is still unclear. Neural information is conveyed jointly by firing rates and spike timing, but existing models treat these dimensions separately, limiting mechanistic interpretation of spike-field coupling and its reported association with spectral features such as the aperiodic slope. Here we show that hippocampal phase locking emerges from a fundamental dynamical mechanism referred to as forced phase integration that separates neural information into orthogonal magnitude ("what") and phase ("when") coordinates. To formalize this principle, the unified complex-valued neuron (UCN) has been developed, a biologically grounded generative framework in which spike timing arises from phase accumulation while spike magnitude encodes instantaneous signal strength. This framework reproduces biological spike-theta synchronization and enables mechanistic re-evaluation of slope-locking associations, demonstrating that previously reported effects arise from oscillatory contamination rather than causal modulation. These findings establish a unified phase-native principle of neural timing and coding.

**1. Introduction:**

The question of how biological neural networks efficiently encode and transmit information remains one of the fundamental challenges in neuroscience. The mammalian hippocampus encodes information using a dual coding strategy in which firing rate and spike timing relative to ongoing network oscillations convey complementary signals. While classical rate-based coding has formed the foundation of modern artificial neural networks (ANNs) [1], converging experimental evidence demonstrates that precise spike timing relative to local field potential (LFP) oscillations carries independent and behaviorally relevant information [2-6]. In hippocampal circuits, theta phase precession enables neurons to encode spatial position within an oscillatory cycle, creating a temporally organized coordinate system that firing rate alone cannot represent [2, 3, 7-9]. This temporal organization is critical for memory formation, navigation, and inter-regional communication [4, 6, 10]. Despite extensive empirical characterization of spike-field coupling (SFC), computational models remain fragmented across incompatible paradigms. Rate-based ANNs effectively encode graded signal magnitude but lack intrinsic temporal dynamics [1, 11]. Standard spiking neural networks (SNNs), including leaky integrate-and-fire (LIF) models, capture discrete spike events but typically reduce output to binary impulses, discarding graded information carried by burst magnitude and subthreshold membrane fluctuations [12, 13]. Phase oscillator and theta-neuron models explicitly represent oscillatory phase and synchronization dynamics [14, 15], yet they operate exclusively in phase space and are not designed to transmit analog signal strength. Statistical point-process and generalized linear models incorporate oscillatory covariates as external regressors rather than treating phase as an internal neuronal state [24, 30]. Thus, to the best knowledge of the authors, no existing framework simultaneously integrates (i) internal phase dynamics, (ii) graded magnitude representation, and (iii) biologically constrained spike generation within a single generative unit.

This theoretical gap becomes particularly evident when interpreting spike-theta phase locking in the presence of broadband, aperiodic neural activity. Neural power spectra are not merely sums of oscillations but include a scale-free $1/f$ component whose exponent reflects network excitation-inhibition (E-I) balance and synaptic timescales [16, 17]. Recent high-impact studies have reported a positive association between the steepness of the aperiodic slope and the strength of theta phase locking in human neurons [18]. However, the mechanistic interpretation of this association remains unresolved. It is unclear whether the aperiodic exponent causally modulates phase locking via altered integration dynamics, or whether it acts as a proxy marker for oscillatory dominance. A major methodological challenge in addressing this question lies in the measurement of phase synchronization itself. Metrics such as the phase locking value (PLV) are biased by spike count and firing rate differences, potentially leading to spurious condition-dependent effects [27]. The pairwise phase consistency (PPC) offers a bias-free estimator of rhythmic synchronization independent of spike count and has therefore become the preferred measure for single-neuron phase locking analysis [27]. However, even unbiased synchronization metrics cannot resolve causality in the absence of a generative model capable of independently manipulating oscillatory forcing and broadband background fluctuations.

To address these limitations, a phase-native computational framework has been formalized in which spike timing emerges from forced phase integration. This principle is instantiated through the unified complex-valued neuron (UCN), a generative neuron model evolving in the complex plane. In this formulation, neural information decomposes into orthogonal magnitude ("what") and phase ("when") coordinates, enabling a principled separation between signal confidence and temporal progression. Unlike threshold-based voltage integration, this phase-native formulation treats timing as a continuous dynamical variable, allowing spike generation to remain synchronized with oscillatory inputs even under heavy noise and frequency drift. By elevating phase to an explicit internal state rather than an emergent byproduct, this framework establishes a mechanistic foundation for spike-field coupling.

We evaluated the proposed framework through three complementary tests spanning theory, simulation, and biological data. First, using controlled synthetic inputs, we show that phase integration alone is sufficient to recover oscillatory structure, operating as a robust phase-locked demodulator, demodulating noisy theta rhythms and outperforming standard LIF models in nonstationary regimes. Second, using mouse hippocampal CA1 recordings during REM and active waking states, we show that phase-native dynamics recapitulate biological spike-theta phase locking with superior fidelity compared to LIF dynamics. Third, we derive a mechanistic explanation of the aperiodic slope-locking association, demonstrating that the apparent correlation arises from shared sensitivity to oscillatory forcing rather than from a direct causal role of the spectral exponent.

Collectively, these results establish the UCN as a unified computational framework capable of bridging rate coding and temporal coding, resolving ambiguities in spectral-locking relationships, and providing a generative, phase-native account of hippocampal spike timing dynamics.

## 2. Theory

This section introduces the conceptual foundations and mathematical formulation of the UCN from various perspectives.

### 2.1 Conceptual Model of Phase-Native Coding

To formalize the dual coding strategy of the mammalian hippocampus, we define the UCN as a dynamical system evolving in the complex plane. Mimicking the biological neurons that exhibit three coupled dynamics: subthreshold oscillation (driven by intrinsic currents and network feedback), phase-locked firing (spiking at specific phases of the rhythm), and variable burst strength (encoding dendritic gain or salience). First, the internal state of a single neuron has been defined utilizing the complex number expressed below:

$$z(t) = r(t)e^{i\theta(t)} \qquad (1)$$

where $z(t)$ denotes the complex internal state of the neuron, the magnitude $r(t) \in R \geq 0$ and the phase $\theta(t) \in R$ represent orthogonal coordinates of information [2, 4]. The magnitude encodes the instantaneous signal intensity, analogous to the firing rate or burst amplitude, while the phase encodes the temporal state of the neuron within an oscillatory cycle, analogous to the somatic membrane potential integration relative to the LFP [3, 4]. This separation allows the neuron to represent "confidence" (via magnitude) and "timing" (via phase) as independent but simultaneous variables, consistent with experimental findings in hippocampal neural cells [3, 4]. Figure 1 graphically depicts the considered computational model for the UCN.

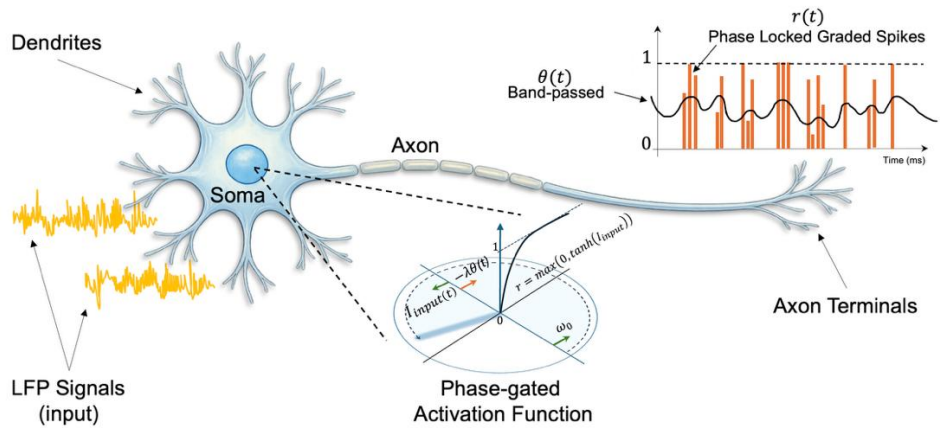

**Figure 1 | Conceptual illustration of the Unified Complex Neuron (UCN).**

Extracellular LFP signals drive the neuron through dendritic inputs. Within the soma, the neuron operates as a phase-gated activation system whose internal state evolves as a forced phase integrator $\theta(t)$. The instantaneous input determines a graded magnitude response $r(t) = max\,(0, tanh\,(Input))$, while the phase variable accumulates until completing a full cycle. A spike event is generated when the internal phase reaches $2\pi$, producing phase-locked graded spikes whose amplitude reflects signal confidence and whose timing reflects oscillatory phase. These spike packets propagate along the axon to downstream neurons, simultaneously transmitting temporal information (phase) and signal magnitude, implementing a unified "what + when" neural coding mechanism.

The schematic illustrates the conceptual interpretation of the UCN. In this framework, each neuron can be modeled as a phase-gated activation unit located in the soma. Extracellular LFP signals are received through the dendrites and drive the internal state of the neuron. The activation function within the soma is governed by an internal phase variable whose dynamics evolve continuously under the influence of these inputs. A spike event is generated when the internal phase completes a full cycle, at which point the neuron emits graded (non-binary) spike packets rather than conventional binary impulses. The amplitude of these spikes reflects the instantaneous confidence or strength of the input signal, while their timing is locked to the phase of the driving oscillation. Consequently, each spike simultaneously conveys both temporal information and signal magnitude, and this information can be transmitted to downstream neurons through the axon terminals.

## 2.2 Mathematical Formulation

### 2.2.1 Magnitude Dynamics and Biological Constraints

The magnitude component $r(t)$ is computed through a weighted aggregation of presynaptic inputs. A critical physiological constraint distinguishes the UCN from standard complex-valued engineering models. In signal processing, it is permissible for amplitudes to be negative (representing a 180° phase shift). However, in a biological context, a "negative spike" is physically undefined; inhibition suppresses firing activity or delays spike timing, but it never inverts the sign of an action potential. To strictly enforce this biological constraint while permitting complex-valued synaptic weights (which facilitate phase-dependent interference), we introduce a rectified activation function. Considering the $Input = \sum_j w_j \, norm(z_j(t))$ as the real-valued projection of the weighted input sum, the internal magnitude is governed by:

$$r(t) = max(0, tanh(Input(t))) \qquad (2)$$

where $w_j$ are complex synaptic weights and $b$ is a bias term. The use of the hyperbolic tangent $tanh(.)$ ensures that the response saturates for strong excitatory inputs, modeling the metabolic limits of neuronal firing. The rectification operator $max(0, tanh(.))$, ensures that strong inhibitory inputs drive the magnitude to zero rather than producing biologically implausible negative amplitudes. Consequently, inhibition in the UCN framework acts as a gating mechanism: it silences the "what" channel (magnitude) while simultaneously retarding the "when" channel (phase), accurately reflecting the subtractive and divisive roles of interneurons [17]. Notice, throughout this research the LFP signals has been considered as driving input for the UCN.

### 2.2.2 Phase Dynamics as a Forced Integrator:

The temporal evolution of the neuron is governed by the phase component $\theta(t)$. We model the phase not as a passive circular variable, but as an active Leaky Integrate-and-Fire mechanism operating in phase space [31, 32]. The dynamics of $\theta(t)$ are described by the following first-order differential equation:

$$\dot{\theta}(t) = -\lambda\theta(t) + \omega_0 + I(t) \tag{3}$$

Here, the $\lambda$ represents a leak factor, ensuring that sub-threshold temporal integration decays over time, mimicking the membrane leak currents in biological neurons and preventing the infinite accumulation of noise [15]. $\omega_0$ represents the intrinsic angular frequency of the neuron, corresponding to its natural tendency to depolarize or "rotate" toward threshold even in the absence of input [31, 32]. The forcing term $I(t)$ represents the aggregate phase drive from presynaptic neurons and external oscillations such as the LFP. This formulation casts the neuron as a forced phase oscillator. Excitatory inputs ($I(t) > 0$, positive values of LFP) accelerate the phase velocity, advancing the spike time (phase precession), while inhibitory inputs ($I(t) < 0$, negative values of LFP) decelerate the phase velocity, delaying the spike time (phase recession). This dynamical structure allows the UCN to inherently demodulate frequency-modulated and phase-modulated signals present in the extracellular environment, providing a mechanistic basis for spike-field coupling [3, 18, 32]

### 2.2.3 Spiking Mechanism and Complex Output Packet

A spiking event is defined as the moment when the accumulated phase completes a full cycle, analogous to threshold crossing in membrane potential models. Specifically, a spike is generated at time $t_{fire}$ when $\theta(t) \geq 2\pi$. As stated before, unlike standard SNNs that emit a binary Dirac delta function $\delta(t)$, Upon crossing this threshold, the neuron emits a complex-valued output packet including a rich information packet $z_{out}(t)$ containing three distinct variables:

$$z_{out}(t) = \delta(t - t_{fire}) \cdot r(t_{fire}) \cdot e^{i\theta(t_{fire})} \tag{4}$$

This output preserves the "Spike-Magnitude-Phase" coding strategy observed in hippocampal pyramidal neurons. On spike events, the downstream network receives the precise timing of the event (phase) and the graded strength of the response (magnitude) simultaneously. Following the event, the internal phase is reset $\theta(t_{fire}^+) = \theta(t_{fire}) - 2\pi$, simulating membrane repolarization [31]. In other words, the UCN framework provides a biophysically interpretable abstraction of hippocampal neurons, filling the representational gaps left by standard computational models. To clearly distinguish the mechanistic and representational advantages of the proposed model, we contrast the UCN with existing rate-based, spiking, and statistical models in Table 1.

**Table 1 | Comparison of representational capabilities and generative mechanisms across neural modeling frameworks.** The Unified Complex Neuron (UCN) is contrasted with established models including Rate-based ANNs [1, 11], Leaky Integrate-and-Fire (LIF) SNNs [12, 31], Phase/Theta models [31, 32], Complex-valued networks [22], and GLM/Point-process frameworks [24, 30]. The UCN is the only framework that simultaneously integrates an internal phase state, graded magnitude representation, and a dynamic temporal coding mechanism.

| Framework | Phase as internal state | Generates spikes | Graded (valued) output | Phase-magnitude interaction | Biologically constrained output | Mechanistic spike-field coupling | Unified "what + when" |
|---|---|---|---|---|---|---|---|
| Rate-based ANNs [1, 11] | No | No | yes | No | Yes | No | No |
| Standard SNNs / LIF models [12, 31] | No | Yes | no | No | Yes | No | No |
| Phase oscillator / theta neurons [31,32] | Yes | Yes | No | No | Yes | Partial | No |
| Complex-valued neural networks [22] | Yes (mathematical) | No | Yes | Yes | No | No | No |
| GLM / point-process models [24,30] | No (external regressor) | Yes | No | No | Yes | No | No |
| UCN | Yes | Yes | Yes | Yes | Yes | Yes | Yes |

## 3. Validation of UCN framework

Presented framework has been evaluated through a two-tier validation strategy combining controlled simulations and in vivo neural recordings.

## 3.1 Datasets and Experimental Design

To rigorously evaluate the proposed phase-native framework, we conducted validation across both controlled synthetic inputs and real biological recordings. This dual-dataset strategy enables mechanistic testing under known ground-truth conditions while assessing biological plausibility under natural neural dynamics.

Synthetic validation employed analytically defined oscillatory signals corrupted by controlled stochastic noise, enabling precise manipulation of frequency stability, phase drift, and signal-to-noise ratio. These experiments isolate the dynamical properties of phase integration and provide a ground-truth environment for benchmarking against threshold-based spiking models.

Biological validation used in vivo hippocampal electrophysiological recordings from the Buzsáki Laboratory public dataset (PeyracheA collection, Mouse17-130125 session), comprising CA1 LFPs and single-unit spike trains recorded from a freely behaving mouse during REM sleep and WAKE states. LFPs were band-pass filtered in the theta range (6-10 Hz), and spike-phase relationships were quantified using pairwise phase consistency and related circular statistics. This dataset provides a biologically realistic testbed for evaluating phase-locking dynamics under natural network oscillations.

## 3.2 Synthetic data validation: UCN as phase demodulator

To rigorously validate the UCN's capacity to extract phase information from biological noise, we subjected the model to a "ground truth" phase-locking experiment. The neuron was driven by a synthetic 8 Hz theta oscillation embedded in heavy additive Gaussian noise ($\sigma = 0.8$), mimicking the low signal-to-noise ratio characteristic of hippocampal LFP recordings (Fig. 2Aa).

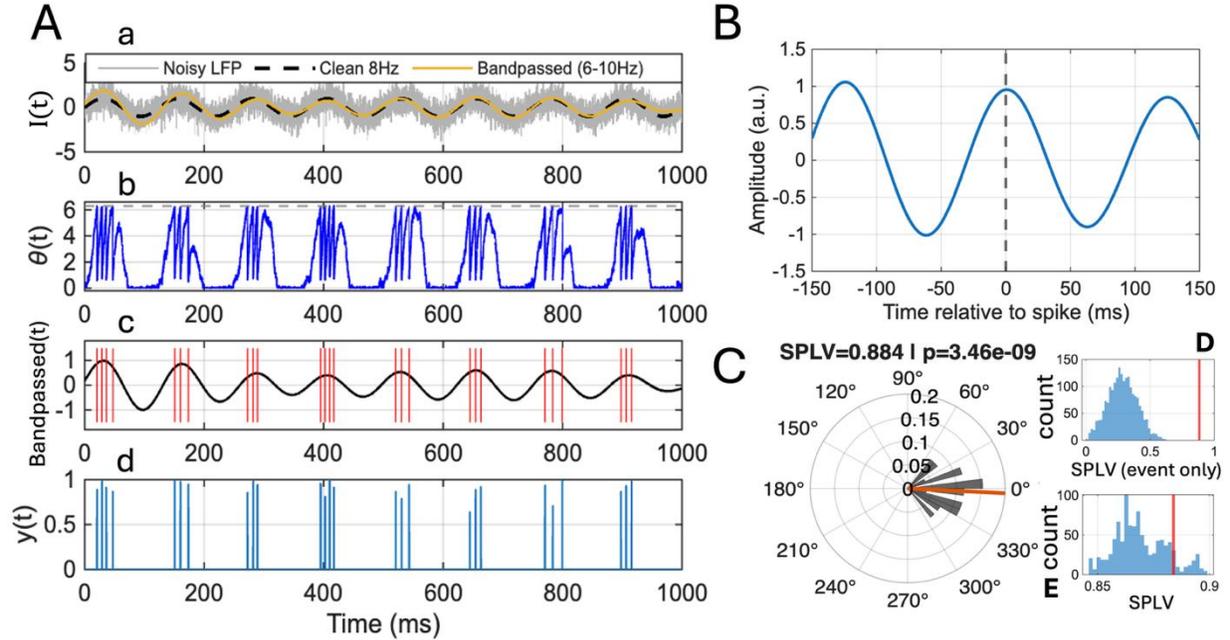

**Figure 2 | UCN acts as a phase-locked demodulator of noisy oscillatory inputs.**

**(A)** Input signal and internal phase dynamics. (a) The synthetic input consists of a clean 8 Hz theta oscillation (dashed black) embedded in heavy Gaussian noise (gray); the band-passed component (6-10 Hz) is shown in orange. (b) The internal phase variable $\theta(t)$ of the UCN integrates the noisy input and resets after reaching $2\pi$, producing characteristic sawtooth dynamics. Variations in slope reflect modulation of phase velocity by instantaneous input strength. (c) Spike events (red lines) relative to the band-passed theta rhythm, illustrating phase-locked firing. (d) Valued output $y(t)$, where spike amplitude reflects the graded magnitude $r(t)$, demonstrating that the model transmits both timing and signal strength. **(B)** Spike-triggered average of the band-passed signal aligned to spike times (vertical dashed line), showing consistent phase alignment of spikes relative to the underlying oscillation. **(C)** Circular histogram of spike phases relative to the 8 Hz oscillation. The UCN exhibits strong unipolar phase locking ($SPLV \approx 0.884$; Rayleigh $p = 3.46e - 9$, indicating that the internal phase integrator effectively filters broadband noise and synchronizes spikes to the oscillatory driver. **(D-E)** Surrogate statistical validation. (D) Jitter surrogate test ($\pm 50$ ms). The observed SPLV (red line) lies far outside the surrogate distribution, confirming that the locking arises from fine temporal structure rather than firing-rate modulation. (E) Matched surrogate test preserving spike statistics, showing that the observed synchronization is consistent with the expected distribution for a phase-locked bursting process.

---

From the perspective of phase demodulation and valued output, despite the significant stochastic noise in the input current, the UCN established a stable phase-locked loop. The internal phase variable $\theta(t)$ (Fig. 2Ab) exhibited characteristic "sawtooth" dynamics, integrating the noisy input until reaching the $2\pi$ threshold. Crucially, the reset mechanism was not rigid; the instantaneous rate of phase progression varied with input strength, allowing the neuron to adjust its firing times to match the external rhythm. The resulting output (Fig. 2Ad) demonstrated the model's dual-coding capability: unlike a standard LIF neuron that emits binary pulses, the UCN generated valued spike packets where the magnitude $r(t)$ varied dynamically with the input's instantaneous confidence, while the spike timing remained precisely locked to the theta trough (Fig. 2Ac). On the other hand, quantitative statistical analysis confirmed robust synchronization. The neuron achieved a SPLV of 0.884 ($N = 26$ spikes), significantly exceeding chance levels (Rayleigh test $z = 21.47, p = 3.46e - 9$). The phase distribution was highly concentrated, with a preferred firing phase of $\mu \approx 356.7°$ (95% CI [346.0°, 7.9°]), indicating alignment near the oscillatory peak/trough boundary (depending on convention) with minimal angular deviation (Fig. 2C). This confirms that the UCN successfully filtered the broadband noise to lock specifically to the 8 Hz carrier frequency. Besides, for further assessments of temporal precision the surrogate testing has been implemented. To prove that this locking arose from precise sub-threshold temporal integration rather than slow firing-rate correlations, we performed a jitter-based surrogate analysis (Fig. 2D-E). We generated 2,000 surrogate spike trains by randomly jittering original spike times within a window of $\pm 50$ ms. This manipulation, which preserves the mean firing rate but destroys fine temporal structure, caused a catastrophic collapse in phase locking (Surrogate SPLV $\mu = 0.273 \pm 0.11$). The observed SPLV of 0.884 fell far outside the surrogate distribution (Empirical $p \approx 0$), confirming that the UCN's performance is driven by a genuine millisecond-scale phase-integration mechanism. Additionally, we compared the results against a "matched" surrogate dataset (circularly shifting spike trains while preserving inter-spike intervals). The UCN's performance remained statistically indistinguishable from the theoretical limit for this burst density (Matched Surrogate $p = 0.789$), suggesting the model captures the maximum available phase information given the sparse firing regime. To establish a baseline for performance, we benchmarked the UCN against a standard Leaky Integrate-and-Fire (LIF) neuron receiving the same noisy theta-modulated input. The LIF model represents the dominant paradigm in SNNs, where input currents are integrated into a scalar membrane voltage V(t) until a threshold is reached.

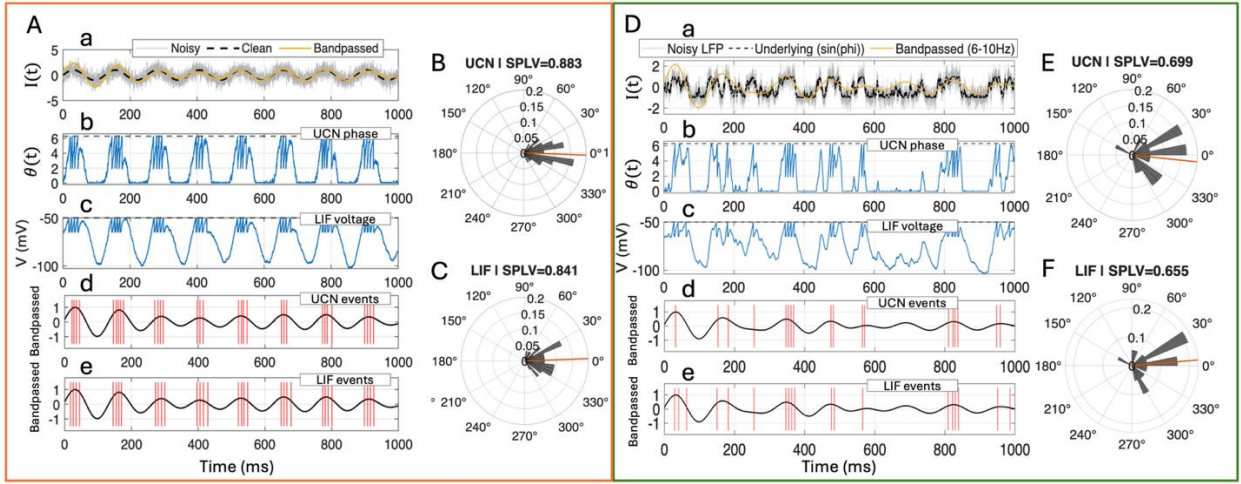

**Figure 3 | Comparative phase-locking behavior of the UCN and LIF under stationary and nonstationary oscillatory inputs**

**(A–C)** Benchmarking under stationary oscillatory input. **(A)** Time-domain dynamics of the input signal, internal state evolution, and spike generation for both models when driven by a noisy oscillatory signal. (a) Synthetic input current $I(t)$ consisting of a clean 8 Hz theta oscillation (dashed black) embedded in strong Gaussian noise (gray), with the band-passed component (6-10 Hz) shown in orange. (b) Internal phase variable $\theta(t)$ of the UCN. The phase integrates the noisy input and resets whenever it reaches $2\pi$, producing the characteristic sawtooth dynamics. Variations in slope reflect modulation of instantaneous phase velocity by the strength of the input signal. (c) Membrane voltage $V(t)$ of the standard LIF neuron driven by the same input current. (d) UCN spike events relative to the band-passed theta rhythm. Spikes occur whenever the internal phase completes a full cycle, producing phase-locked spike packets. (e) LIF spike events generated when the membrane voltage crosses threshold. **(B)** Circular histogram of spike phases for the UCN relative to the underlying oscillatory driver. The model exhibits strong unipolar phase locking with a $SPLV$ of 0.883. The resultant vector (orange) indicates a strong preferred firing phase close to the oscillatory peak/trough boundary. **(C)** Phase distribution of spike events for the LIF model under identical conditions. Although the LIF neuron also synchronizes to the oscillatory input ($SPLV = 0.841$), the angular concentration is weaker compared to the UCN, indicating reduced synchronization precision. **(D–F)** Robustness under nonstationary oscillatory input with phase random-walk. **(D)** Time-domain dynamics when the input oscillation exhibits stochastic phase drift. (a) Input signal generated with a random-walk phase process ($\sigma = 10 \, rad\sqrt{s}$) to mimic the nonstationary nature of biological oscillations. The gray trace shows the noisy signal, the dashed line indicates the underlying drifting oscillation, and the orange trace shows the band-passed component. (b) Internal phase state of the UCN tracking the drifting oscillation through continuous phase integration. (c) Membrane voltage dynamics of the LIF neuron under the same input conditions. (d) Spike events generated by the UCN relative to the band-passed oscillation. (e) Spike events generated by the LIF neuron. **(E)** Circular histogram of spike phases for the UCN under nonstationary conditions. Despite the stochastic frequency drift of the driving oscillation, the UCN maintains stable synchronization with $SPLV = 0.699$. **(F)** Phase distribution for the LIF neuron under the same drifting input. The LIF model exhibits weaker phase concentration ($SPLV = 0.655$), reflecting increased phase slips caused by threshold-based integration when the oscillation frequency varies.

---

Results shows that the LIF model successfully established phase locking to the 8 Hz driver with SPLV = 0.841 and N = 31 (Fig. 3C), confirming that standard voltage integration is a satisfactory mechanism for basic temporal synchronization whilst The UCN matched and slightly exceeded this performance achieving SPLV = 0.883 and N = 32 under identical noise conditions (Fig. 3B). This confirms that the UCN framework preserves the fundamental "satisfactory" timing capabilities of the LIF model; it does not sacrifice temporal precision to achieve its complex formulation. Thus, it can be inferred like that the UCN is not sacrificing anything to provide a richer informational output. Here, the results can confirm while both models satisfactorily capture spike timing (when information), they diverge fundamentally in what information is transmitted. The LIF neuron reduces the rich, continuous structure of the LFP into a binary event series (0 or 1), effectively discarding all information regarding the instantaneous strength or "confidence" of the LFP signal at the moment of firing. In contrast, the UCN provides a lossless demodulation of the input. This allows the downstream network to receive the same precise timing information as the LIF (the "When") while simultaneously recovering the graded signal intensity (the "What") that the LIF model ignores. Thus, the UCN does not just mimic the standard spiking behavior; it enriches it, offering a more complete readout of the underlying neural state without requiring additional transmission channels. To rigorously distinguish the tracking capabilities of the two models under biologically realistic conditions, we subjected both to a nonstationary phase challenge. In biological circuits, theta oscillations are rarely stable sinusoids; they exhibit frequency drift and phase slips (random walks). We modeled this by driving both neurons with a "phase random walk" signal ($\sigma_\phi = 10.0\ rad/\sqrt{s}$), introducing significant unpredictability into the cycle timing (Fig. 3Da). While both models maintained significant locking (Rayleigh $p < 0.001$), the UCN demonstrated superior tracking of the drifting rhythm. The UCN achieved an $SPLV$ of 0.699 and $N = 18$, consistently outperforming the LIF model with $SPLV = 0.655$ and $N = 19$ under identical nonstationary conditions. The temporal dynamics reveal the mechanism: the UCN's equation $\dot{\theta} = -\lambda\theta + \omega_0 + I(t)$ allows the input current to directly modulate the instantaneous phase velocity. When the external rhythm accelerates or decelerates (frequency drift), the UCN adapts its rotation speed continuously. In contrast, the LIF neuron must integrate a fixed charge threshold to fire; if the rhythm accelerates, the integration window shortens, causing the LIF to "slip" or miss cycles more frequently.

Collectively, these simulations establish that the UCN functions as a robust phase demodulator, maintaining precise locking ($p \approx 0$) even under heavy noise (SNR < 0 dB). Crucially, the model outperforms standard LIF dynamics in nonstationary regimes: while both models track stable rhythms, the UCN's direct phase-integration mechanism allows it to continuously adapt to frequency drifts and phase slips that cause threshold-based models to decouple. By combining this superior tracking stability with a valued output that preserves signal magnitude, the UCN provides a richer and more biologically plausible readout of oscillatory information than binary spiking models. To further evaluate the robustness of the phase-integration mechanism under nonstationary conditions, we subjected both models to oscillatory inputs with stochastic phase diffusion, mimicking the frequency drift commonly observed in biological theta rhythms. In this experiment, the phase of the driving oscillation followed a diffusion process controlled by the parameter $\sigma_\phi$, which progressively destabilizes the rhythm and introduces random phase slips. Increasing $\sigma_\phi$ therefore creates a challenging synchronization regime resembling the frequency drift commonly observed in biological theta oscillations.

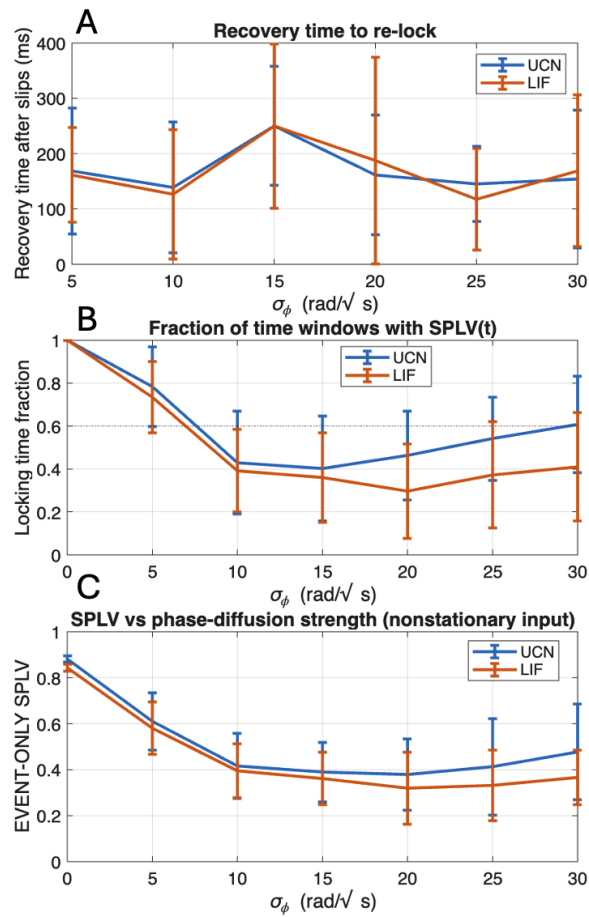

**Figure 4 | Robust phase tracking of nonstationary oscillations by the UCN**

Performance of the UCN compared with a standard LIF neuron under progressively increasing phase diffusion in the driving oscillatory input. Phase diffusion strength is controlled by the parameter $\sigma_\phi$ ($rad/\sqrt{s}$), which introduces stochastic phase drift mimicking the nonstationary dynamics commonly observed in biological theta oscillations. Results are averaged over 30 trials per condition. **A**) Recovery time following phase slips. The average time required for each neuron model to re-establish phase locking after a phase slip event is shown as a function of diffusion strength. Both models exhibit longer recovery times as phase diffusion increases; however, the UCN maintains comparable or slightly faster re-locking dynamics with reduced variability across trials. **B**) Fraction of time windows exhibiting significant phase locking. The proportion of time windows in which spike trains remain phase locked to the oscillatory input is shown for each model. As phase diffusion increases, both models experience reduced locking stability, but the UCN consistently maintains a larger fraction of locked windows compared with the LIF neuron. **C**) Event-only SPLV as a function of phase diffusion strength. Increasing phase diffusion progressively degrades synchronization for both models; however, the UCN consistently achieves higher SPLV values than the LIF baseline across all noise regimes, indicating more robust tracking of drifting oscillatory inputs.

Error bars represent ±1 standard deviation across trials. Collectively, these results demonstrate that the phase-integrator dynamics of the UCN enable more stable synchronization with nonstationary oscillations compared with threshold-based voltage integration in conventional spiking neuron models.

As shown in Fig. 4, both neuron models exhibit gradual degradation of synchronization as diffusion strength increases, but their dynamical responses differ substantially. The UCN consistently recovers from phase disruptions with stable re-locking dynamics (Fig. 4A), whereas the LIF model shows larger variability in recovery time across trials. This difference becomes more apparent when examining the temporal stability of synchronization: the UCN remains phase locked for a larger fraction of time windows across the entire diffusion range (Fig. 4B), indicating greater resilience to phase perturbations. A similar trend is observed in the spike-phase synchronization statistics. Although increasing phase diffusion reduces the spike-phase locking value for both models, the UCN systematically maintains stronger locking compared with the LIF baseline (Fig. 4C). Notably, the performance gap becomes more pronounced at higher diffusion levels, where the oscillatory phase undergoes stronger stochastic drift. These results indicate that directly integrating phase dynamics enables the UCN to track drifting oscillatory inputs more reliably than threshold-based voltage integration, supporting the interpretation of the UCN as a phase-native demodulation mechanism capable of maintaining synchronization under biologically realistic nonstationary conditions.

These controlled simulations demonstrate that phase-native integration enables stable tracking of nonstationary oscillations, providing a mechanistic basis for evaluating the UCN framework on real hippocampal electrophysiological recordings.

### 3.3 Biological data Validation: Demodulating hippocampal theta rhythm

Having established the UCN as a robust phase demodulator in silico, we next challenged the model with real electrophysiological data to determine its biological plausibility. We utilized Local Field Potential (LFP) recordings from the mouse hippocampus (CA1) during two distinct behavioral states: Rapid Eye Movement (REM) sleep, characterized by stable, high-amplitude theta oscillations; and Active Waking (WAKE), characterized by more transient and modulated theta rhythms associated with exploration. This dual-regime testing provides a rigorous "ground truth" environment to determine if the UCN can autonomously synchronize with biological rhythms derived from neural circuits rather than mathematical functions.

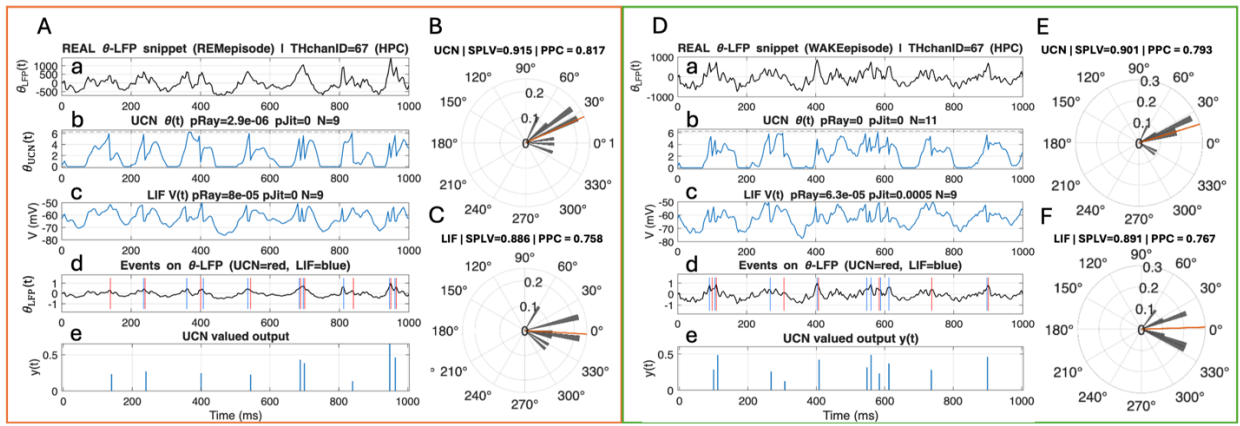

**Figure 5 | UCN acts as a digital twin for hippocampal theta phase dynamics in real CA1 LFP during REM and WAKE episodes.** Real LFP was recorded from the hippocampal theta channel (THchanID=67; region tag: HPC; fs=250 Hz). For each behavioral state, a 1 s snippet was used to drive both the UCN and a rate-matched LIF baseline, and spike-theta synchronization was quantified using event-only spike phases. **(A–C)** REM episode, **(A)** Example REM theta-LFP snippet and model dynamics. (a) Raw $\theta_{LFP}(t)$ segment. (b) UCN internal phase state $\theta_{UCN}(t)$, showing sawtooth-like phase accumulation and reset at $2\pi$. (c) LIF membrane voltage $V(t)$ driven by the same LFP as input. (d) Detected spike events overlaid on the theta-LFP phase trace (UCN events in red; LIF events in blue), illustrating alignment of spiking to a preferred theta phase. (e) UCN valued output $y(t)$, showing graded spike packets whose amplitude reflects instantaneous input-driven magnitude. **(B)** Circular histogram of UCN

spike phases during REM, demonstrating strong phase locking ($SPLV = 0.915$, $N = 9$; Rayleigh $z = 7.532$, $p = 2.89e - 6$; preferred phase $\mu = 24.1°$, 95% $CI$ [8.2°, 39.7°]; $PPC = 0.817$). **(C)** Corresponding LIF spike-phase distribution during REM ($SPLV = 0.886$, $N = 9$; Rayleigh $z = 7.064$, $p = 8.01e - 5$; preferred phase $\mu = 356.6°$, 95% $CI$ [340.2°, 16.6°]; $PPC = 0.758$). The UCN exhibits higher synchronization fidelity than LIF in REM ($\Delta SPLV = 0.0289$; $\Delta PPC = 0.0585$). **(D–F)** WAKE episode, **(D)** Example WAKE theta-LFP snippet and model dynamics, shown with the same layout as in (A). UCN and LIF spike events are overlaid on the theta-LFP phase trace (UCN red; LIF blue), and UCN valued output $y(t)$ is shown in (e). **(E)** UCN spike-phase distribution during WAKE ($SPLV = 0.901$, $N = 11$; Rayleigh $z = 8.934$, $p \approx 0$; $\mu = 17.2°$, 95% $CI$ [0.9°, 31.6°]; $PPC = 0.7934$). **(F)** LIF spike-phase distribution during WAKE ($SPLV = 0.8905$, $N = 9$; Rayleigh $z = 7.137$, $p = 6.28e - 5$; $\mu = 1.9°$, 95% $CI$ [345.8°, 20.7°]; $PPC = 0.767$). The UCN maintains a consistent advantage over LIF during WAKE ($\Delta SPLV = 0.0107$; $\Delta PPC = 0.0263$).

The UCN was first driven by raw, normalized LFP signals recorded from the pyramidal layer of the CA1 hippocampus during REM sleep episodes (Fig. 5Aa). Crucially, the model was not provided with any explicit information regarding the LFP frequency or phase; it was required to extract this information solely through its internal differential equations. The UCN successfully entrained to the biological theta rhythm, exhibiting precise phase-locked firing analogous to the behavior of native hippocampal pyramidal neurons. During REM episodes, the model achieved a $SPLV$ of 0.915 and a $PPC$ of 0.817 ($N = 9$ spikes, Rayleigh $p < 10^{-5}$), demonstrating that the internal phase variable $\theta(t)$ effectively tracked the physiological oscillation (Fig. 1Aa). The preferred firing phase was tightly clustered near the theta trough ($\mu \approx 24°$), consistent with experimental observations showing that hippocampal pyramidal neurons exhibit preferred theta-phase firing that strengthens during high-theta states and enhanced rhythmic engagement [3, 4, 25]. To validate that this synchronization was not a stochastic artifact, we performed a jitter-based surrogate analysis ($\pm 50$ ms), which resulted in a catastrophic loss of locking (Surrogate $SPLV \approx 0.3$, Empirical $p \approx 0$), confirming that the UCN's response is driven by millisecond-scale temporal structure within the LFP. Observations confirm the tracking precision of UCN in REM theta dynamics with an acceptable performance. Further, to assess the robustness of the model across behavioral states, we repeated the analysis on LFP segments recorded during WAKE episode. The WAKE theta is often more nonstationary than REM theta, presenting a greater challenge for phase tracking. The UCN maintained robust synchronization in this regime, achieving an $SPLV$ of 0.901 and a $PPC$ of 0.793 (Fig. 5E). The preferred phase shifted slightly ($\mu \approx 17.2°$) but remained stable within the theta trough, mirroring the phase stability often observed in biological place cells across states. This confirms that the UCN's phase-integration mechanism is not reliant on the "clean" sinusoidal nature of REM theta, but it is robust enough to demodulate the more complex, broadband oscillatory activity associated with active behavior.

Now, a key question for the UCN framework is whether its complex-valued phase integrator provides a measurable advantage over standard real-valued LIF dynamics when processing biological signals. We subjected a standard LIF neuron to the same hippocampal LFP inputs, tuning its parameters to match the UCN's firing rate. While the LIF neuron also established phase locking, the UCN demonstrated consistently superior synchronization fidelity across both behavioral states. During REM sleep, the UCN achieved a significantly higher pairwise phase consistency with $\Delta PPC = +0.059$ compared to the LIF model (Fig. 5B-C). This advantage persisted during WAKE episodes with $\Delta PPC = +0.026$ (Fig. 5E-F). This performance gap highlights a fundamental mechanistic difference: the LIF neuron integrates voltage and fires only when a threshold is crossed, a process prone to "slipping" when the biological theta amplitude fluctuates. In contrast, the UCN integrates phase velocity directly ($\dot{\theta} = -\lambda \theta(t) + \omega_0 + I(t)$), allowing it to maintain the phase progression even during transient dips in LFP amplitude that cause standard threshold models to miss cycles. Beyond timing precision, the UCN provided a richer informational readout than the binary LIF model. As observed in the real-data implementations (Fig. 5Ae, and Fig. 5De), the UCN emitted valued spike packets where the magnitude $y(t)$ varied dynamically with the instantaneous strength of the LFP envelope for example the value of 8[th] spike emitted in REM episode is highly larger than the value of the first spike that can convey that in firing time for 8[th] spike the LFP signal was stronger than the LFP signal at the firing time of the first spike (Fig. 5Ae), and such information will not be extractable with LIF. These outputs structure for UCN effectively creates a digital replica of the "Spike-Magnitude-Phase" code, where the timing of the output spike reflects the LFP phase, and the magnitude of the output packet reflects the instantaneous excitability or "confidence" of the circuit. This confirms that the UCN is capable of acting as a neural codec for biological signals, preserving the multi-dimensional information (What + When) inherent in hippocampal processing.

## 5. Mechanistic Analysis of Aperiodic Slope Effects

To provide a rigorous mechanistic explanation for the empirical correlation between the estimated aperiodic slope and theta phase locking reported by Guth et al. [18], we formally derive the phase dynamics of the UCN under conditions of oscillatory forcing. We begin by defining the phenomenological observation where $\hat{\beta}$ represents the estimated aperiodic spectral exponent derived from a standard log-log fit of the local field potential power spectrum, and $PPC$ denotes the pairwise phase consistency of spiking activity relative to the theta rhythm. The reported findings correspond to a non-zero covariance between these variables, $Cov(\hat{\beta}, PPC) \neq 0$, specifically indicating that "steeper" slopes (larger $|\hat{\beta}|$ in log-log coordinates) are associated with stronger phase locking. In the UCN framework, spike timing is generated by an internal phase state $\theta(t)$ that integrates a leak, intrinsic drive, and input forcing. In the continuous-time approximation of the dynamics, the phase evolution is given by the differential equation expressed in Eq. (3) with $I(t) = kx(t)$, where $x(t)$ is the normalized LFP-driven input. The input $x(t)$ can be decomposed into a coherent oscillatory theta component and a broadband aperiodic background term, expressed as $x(t) = A_\theta \cos(\omega_0 t + \psi) + \xi(t)$. Substituting this decomposition into the Eq. (3) yields the forced Langevin dynamics:

$$\dot{\theta} = -\lambda\theta(t) + \omega_0 + kA_\theta \cos(\omega_0 t + \psi) + k\xi(t) \tag{5}$$

On the other hand, spike generation is governed by a threshold-and-wrap mechanism where a spike event occurs at $t_{fire}$ if $\theta(t_{fire}) \geq 2\pi$, followed by the reset $\theta \leftarrow \theta - \alpha(2\pi)$. Thus, phase locking of spike times to the theta phase is an emergent property of a periodically forced leaky integrator.

## 5.1 Analytical analysis

Under standard phase-reduction intuition, the effect of broadband fluctuations $\xi(t)$ appears as an effective phase diffusion term. The phase deviation $\varphi = \theta_{UCN} - \theta_{LFP}$ relative to the forcing evolves according to the stochastic differential equation $d\varphi = -\kappa \sin(\varphi)dt + \sqrt{2D}dW_t$. Here, $\kappa$ is a forcing-dependent stability term that increases with the coupling strength $kA_\theta$, and $D$ is an effective diffusion constant induced by the broadband term $k\xi(t)$. The concentration of spike phases, and therefore the $PPC$, is a monotonic function of the ratio $\kappa/D$. This derivation highlights two critical dependencies: $PPC$ is primarily controlled by the theta forcing strength $A_\theta$ (via $\kappa$) which means the tighter phase locking accompanies stronger theta oscillations, while the broadband background affects locking mainly through its effective diffusion magnitude $D$, rather than the spectral exponent alone. On the other hand, the correlation between slope and locking arises from the estimation of $\hat{\beta}$. The LFP power spectrum is a mixture $S(f) = Cf^{-\beta_{true}} + G(f; A_\theta, f_\theta)$, where the first term is the true aperiodic component and $G(.)$ captures the oscillatory theta peak. When $\hat{\beta}$ is estimated by a linear fit in log-log space over a band that includes the theta frequency, the oscillatory term introduces an estimation bias, such that $\hat{\beta} \approx \beta_{true} + g(A_\theta)$. Since the theta amplitude $A_\theta$ physically strengthens locking in the UCN via $\kappa \propto A_\theta$, both the estimated slope $\hat{\beta}$ and the PPC increase simultaneously due to the shared latent driver $A_\theta$. Consequently, observing $Cov(\hat{\beta}, PPC) \neq 0$ does not imply that the aperiodic exponent causally determines phase locking.

## 5.2 Statistical Analysis

We empirically validated this theoretical framework by analyzing the UCN's performance on real mouse hippocampal LFP data under two distinct analytical regimes.

### 5.2.1 Aperiodic Slope Estimation without Oscillatory Peak Removal

In the first regime, we replicated the standard analytical pipeline by estimating the aperiodic slope $\hat{\beta}$ over a broad frequency band (1-40 Hz) that included the theta oscillation (6-10 Hz). Under these conditions, which are susceptible to oscillatory contamination, the UCN faithfully reproduced the literature. In the REM sleep state, where theta oscillations are most prominent, epochs classified as having steep slopes were associated with significantly higher phase locking of $Mean\ PPC_{steep} = 0.7902$ compared to epochs with flat slopes with $Mean\ PPC_{flat} = 0.7639$. This difference was highly statistically significant with $p = 5.87 \times 10^{-6}$, confirming that the model captures the strong empirical association reported in the literature. A similar significant trend was observed in the WAKE state, where steep slopes with $PPC = 0.735$ yielded higher locking than flat slopes $PPC = 0.7195; p = 0.027$. These results confirm that $\Delta PPC > 0$ when the confounding variable $A_\theta$ is allowed to influence the slope estimate.

### 5.2.2 Aperiodic Slope Estimation with Oscillatory Peak Removal

To isolate the true causal driver, we applied a rigorous spectral control. We re-estimated the aperiodic slope by explicitly excluding the theta frequency band (6-10 Hz) from the spectral fit and enforcing temporal separation between analysis windows to ensure statistical independence. Under these controlled conditions, the association between slope and phase locking weakened substantially or vanished. In the active waking state (WAKE), the difference in phase consistency between steep ($PPC = 0.7380$) and flat ($PPC = 0.7398$) slope conditions became statistically indistinguishable from zero ($\Delta Mean = -0.0018$; Permutation test $p = 0.935$). Similarly, in the REM state, the effect size diminished ($\Delta Mean = 0.0237$) and lost statistical significance ($p = 0.240$).

Results here, provide definitive support for the mechanistic interpretation derived in Section 6.1. When the oscillatory bias is removed ($\frac{\partial}{\partial \beta_{true}} E[PPC|A_\theta] \approx 0$), the residual dependence of phase locking on the aperiodic exponent is negligible. This confirms that phase locking is primarily governed by the strength of theta-band oscillatory forcing $A_\theta$, while the apparent association between spectral slope and locking arises from how the estimated aperiodic exponent $\hat{\beta}$ is influenced by oscillatory power during slope estimation. In this sense, previously reported slope-locking correlations reflect a shared sensitivity to oscillatory dominance rather than a direct causal role of the aperiodic exponent itself. The UCN model resolves this apparent paradox by identifying oscillatory dominance as the common driver underlying both steeper estimated slopes and tighter spike-theta phase synchronization. Thus, the aperiodic slope functions as a state indicator rather than a control parameter for phase locking. Our results reconcile prior observations by showing how oscillatory dominance can simultaneously influence spectral slope estimation and spike-theta locking, without requiring a direct causal role for the aperiodic exponent. Table 2 summarizes the analyzes conducted here.

**Table 2 | Oscillatory control removes the apparent slope–locking association.**

Steeper slopes are associated with stronger phase locking when theta activity is included in slope estimation. This association weakens or disappears after oscillatory contamination is removed, indicating a shared sensitivity to oscillatory dominance rather than a direct causal role of the aperiodic exponent.

| Statistical Regime | Brain State | Slope–Locking Association | Statistical Outcome |
| --- | --- | --- | --- |
| Uncontrolled Estimation | WAKE | Steep slopes show stronger locking | Significant |
|  | REM | Steep slopes show markedly stronger locking | Highly significant |
| Controlled Estimation | WAKE | No meaningful difference between slope groups | Not significant |
|  | REM | Weak residual difference | Not significant |

## 6. Discussion

This study establishes a phase-native computational account of hippocampal spike timing by showing that spike-theta phase locking emerges directly from forced phase-integration dynamics. Across controlled synthetic conditions and real mouse hippocampal recordings, biologically realistic spike-field coupling arises intrinsically from oscillatory phase evolution without requiring explicit phase alignment or heuristic post-processing. These findings indicate that neural phase locking can be understood as a natural consequence of dynamical integration of rhythmic inputs rather than as an externally imposed synchronization phenomenon.

This perspective reframes how temporal coordination is represented in neural systems. In conventional rate-based neural networks, temporal structure is encoded implicitly through firing rates or population averages [1, 2]. Classical spiking neuron models can exhibit phase locking, but synchronization typically emerges indirectly from membrane-voltage dynamics and often depends on careful parameter tuning or external phase-referencing mechanisms [12, 31]. Signal-processing approaches characterize phase relationships descriptively yet do not provide a generative account of how phase-locked spiking arises from neuronal dynamics [23, 24]. By contrast, a phase-native formulation treats oscillatory phase as an explicit dynamical state variable that directly governs spike timing. Embedding oscillatory integration at the core of neuronal dynamics reveals spike timing as an intrinsic outcome of phase evolution rather than an externally imposed feature. This mechanistic view aligns with extensive evidence that neural oscillations coordinate spike timing to enable communication and computation across distributed brain networks [3-6, 25].

A central conceptual advance of this work is the mechanistic resolution of the relationship between aperiodic spectral slope and theta phase locking. Prior studies have reported correlations between steeper spectral slopes and stronger spike-field synchronization [16-18], suggesting that broadband spectral structure may modulate neural timing. The present analysis shows that this association does not imply a direct causal role of the aperiodic exponent. Instead, both effects arise from a shared latent driver: oscillatory dominance. This interpretation is consistent with prior reports that stronger theta rhythmic engagement is associated with tighter spike-theta phase locking in hippocampal neurons [25]. When oscillatory power contributes to slope estimation, variations in theta forcing strength simultaneously bias the estimated aperiodic slope and enhance spike-theta synchronization. Removing oscillatory contamination eliminates residual slope dependence, demonstrating that phase locking is governed primarily by oscillatory drive rather than by the aperiodic component itself. This result reframes slope-locking correlations by distinguishing statistical association from mechanistic causation and explains previously inconsistent findings across analytical pipelines.

More broadly, these findings reveal how spectral-domain descriptors and time-domain synchronization can be unified within a single dynamical framework. Spectral parameterization choices directly influence downstream physiological inference, underscoring the need for oscillation-aware analysis when linking spectral features to neuronal timing phenomena [16, 23]. A phase-native dynamical formulation bridges this divide, showing that both spectral structure and spike synchronization emerge from shared phase-reduction and synchronization principles [31, 32].

Conceptually, treating phase as an explicit internal state variable rather than an emergent byproduct distinguishes this framework from conventional rate-based networks and classical spiking neuron models. This formulation enables a principled separation between temporal coordination and activation magnitude, closely aligning with foundational concepts in systems neuroscience including spike–field coupling, phase coding, and oscillation-mediated communication [3-6, 25]. As a result, neural timing becomes generative, interpretable, and analytically tractable within nonlinear dynamical systems theory [31, 32].

Beyond resolving a specific empirical ambiguity, these results establish a foundation for phase-aware neural computation. Operating directly in the representational domain used by oscillatory brain activity enables more faithful decoding of neural dynamics, improved interpretation of state-dependent modulation, and principled linkage between spectral features and neuronal mechanisms. Extending this framework toward closed-loop neural systems may enable physiologically aligned readout and controlled phase-aware stimulation of biological neural tissue, including organoids, while minimizing maladaptive synchronization. In the longer term, phase-native neural codecs could support bidirectional brain–machine interface architectures for restoring or modulating disrupted neural coordination, with potential relevance to neurological disorders such as Parkinson's and Alzheimer's disease.

## 7. Conclusion

This study establishes a phase-native computational account of hippocampal spike timing by showing that spike-theta phase locking emerges from forced phase-integration dynamics across both synthetic and real neural recordings. By treating phase as an explicit internal dynamical variable, this framework reproduces biologically observed phase-locking statistics and enables a principled dissociation between oscillatory timing and activation strength. Through theoretical analysis and controlled statistical evaluation, we demonstrate that theta phase locking is governed primarily by oscillatory forcing, whereas the reported association with aperiodic spectral slope arises from estimation bias introduced by oscillatory power.

These findings reconcile prior observations by identifying oscillatory dominance as the common driver underlying both steeper estimated slopes and tighter spike-theta synchronization, without requiring a direct causal role for the aperiodic exponent. Accordingly, the aperiodic slope is more appropriately interpreted as an indicator of network state rather than a control parameter for neural timing.

Beyond resolving this specific ambiguity, the results establish a foundation for phase-native models of neural computation and for more faithful decoding and interpretation of brain signals structured by network oscillations.

## 8. Data Availability

Electrophysiological recordings are publicly available from the Buzsáki Laboratory dataset (PeyracheA collection, Mouse17–130125): https://buzsakilab.nyumc.org/datasets/PeyracheA/Mouse17/Mouse17-130125/

## 9. Code Availability

A minimal, reproducible version of the code supporting the findings of this study is publicly available at: *https://github.com/INQUIRELAB/Phase-native-neural-computing-/tree/main* Additional components are available from the corresponding author upon reasonable request for academic use. Full implementation details are not disclosed due to an ongoing United States patent application covering key elements of the UCN framework.